\documentclass[aps,pre,twocolumn,reprint,superscriptaddress,nofootinbib]{revtex4-2}

\usepackage[utf8]{inputenc}
\usepackage[T1]{fontenc}
\usepackage[english]{babel}
\usepackage{amsmath,amssymb}
\usepackage{graphicx}
\usepackage{booktabs}
\usepackage{siunitx}
\usepackage{tikz}
\usepackage{pgfplots}
\pgfplotsset{compat=1.17}
\usetikzlibrary{arrows.meta,positioning,calc,decorations.pathreplacing,patterns}
\usepackage[colorlinks=true,allcolors=blue]{hyperref}

\graphicspath{{figures/}}

\newcommand{\rhob}{\bar\rho}
\newcommand{\ton}{t_{\mathrm{on}}}

\definecolor{mplC0}{HTML}{1F77B4}
\definecolor{mplC1}{HTML}{FF7F0E}
\definecolor{mplC2}{HTML}{2CA02C}
\definecolor{mplC3}{HTML}{D62728}
\definecolor{seisB}{HTML}{0000C8}
\definecolor{seisR}{HTML}{B40000}

\begin{document}

\title{Self-gravitating dust clumps: linear theory sets the threshold, the onset time and the spacing of their formation, then slow coarsening erases this pattern}

\author{Yohann Trivino}

\begin{abstract}
Planets form in disks of gas and dust. One step of their formation is the gathering of dust grains into dense clumps, under their own gravity. This work tries to predict when these clumps appear and how far apart they are. The grains attract each other through a gravity of finite range. The gas slows them down, and its turbulence spreads them out. The grains also cannot go beyond maximum packing: this is volume filling. The dimensionless model depends on two numbers. The first one is the gravitational number $\chi$, which compares attraction with turbulent spreading. The second one is the mean volume fraction $\rhob$, which measures the amount of grains. We solve the model with second-order finite volumes on an adaptive Cartesian mesh. We then treat each field as an image. An Otsu threshold and a watershed transform isolate the clumps. The code conserves mass to within \num{e-15} and converges at second order. Two designs of experiments give 192 simulations in total, with $2\le\chi\le40$ and $0.05\le\rhob\le0.7$. Linear stability theory predicts the start of clump formation. The instability starts at time $\ton\simeq3.0/\sigma^*$, where $\sigma^*$ is the growth rate of the most amplified mode. The dominant wavelength equals the most amplified wavelength to within \SI{10}{\percent}. The area fraction of the clumps equals $\rhob$. The clumps then go through slow coarsening. A sensitivity analysis attributes most of the variance to $\chi$. The adaptive mesh reduces the computational cost only a little, because the clumps fill the whole domain.
\end{abstract}

\maketitle

\section{Introduction}

Planets form in disks of gas and dust around young stars. At the beginning, the dust is made of micrometer-sized grains. Blum and Wurm showed that these grains first grow by sticking together when they collide~\cite{blum2008}. Above one millimeter, however, collisions make the grains bounce or break more often than they make them stick. Collisions alone therefore do not lead to planetesimals, the kilometer-sized bodies that come before planets. Another route is collective. Many grains gather into dense clumps, which then collapse under their own gravity~\cite{chiang2010}. This paper studies the formation of these clumps.

Each grain attracts the other grains through gravity. The gas slows the grains down, so they do not accelerate. Weidenschilling described this drag in disks~\cite{weidenschilling1977}: a grain under a force very quickly reaches a constant velocity, proportional to this force. The grains therefore slide at constant velocity toward the bottom of the gravitational potential wells. The turbulence of the gas also shakes the grains randomly. Youdin and Lithwick showed that this shaking spreads the grains like a diffusion~\cite{youdin2007}. The effect of gravity amplifies itself. A slightly denser region digs a deeper potential well. It therefore attracts more grains and becomes even denser. In this way, a dust layer that is uniform at first breaks up into separate clumps. Figure~\ref{fig:mecanisme} illustrates this mechanism.

However, the grains have a volume. They cannot be packed beyond maximum packing, like marbles in a box. This is called volume filling. Without this limit, gravity would concentrate the whole mass of a clump into one point, in a finite time~\cite{chavanis2008}. With this limit, the clumps become plateaus of packed grains. Their number then decreases very slowly, through successive mergers and disappearances. This process is called coarsening. This paper measures its speed.

The equations of this problem have a useful property. Apart from the names of the variables, they are identical to the Keller--Segel model~\cite{keller1970}. This model originally describes amoebae that gather by following a chemical signal. It dates from 1970, but it is still widely studied~\cite{horstmann2003,hillen2009}. Chavanis showed the equivalence between the two problems in detail~\cite{chavanis2008}. Painter and Hillen added volume filling to the model~\cite{painter2002}. Dolak and Schmeiser, and then Potapov and Hillen, studied its coarsening~\cite{dolak2005,potapov2005}. All these results therefore apply directly to dust grains.

The model strongly simplifies the physics of a disk. First, it replaces the rotation of the disk by a gravity of finite range. Second, it neglects the effect of the grains on the gas. This effect creates another instability, the streaming instability~\cite{youdin2005}. This dynamics is rich, but it does not change the question asked here. The model therefore keeps only three ingredients: gravitational attraction, turbulent spreading and the crowding of grains.

This work extends two previous studies. The first one simulates, grain by grain, the gravitational collapse of a dust cloud without gas~\cite{trivino2025}. The second one simulates the collision of two aggregates of bonded grains~\cite{trivino2026}. The present work takes a complementary point of view. It describes the grains by their density, at a scale where many clumps form at the same time.

The goal of this work is to predict two simple quantities. The first one is the time at which the clumps appear. The second one is the distance between them. Linear stability gives a prediction for each of them. It describes the growth of very small irregularities in the uniform distribution of the grains. Keller and Segel applied it to their model from the start~\cite{keller1970}, and Painter and Hillen applied it to the volume-filling version~\cite{painter2002}. In principle, this theory is valid only as long as the irregularities stay small. But a clump is a large irregularity. The question is therefore the following. Does linear theory still predict the onset time and the spacing of the clumps, once they are fully formed?

A second question concerns the measurement. Disk simulations produce maps of dust density. For example, Johansen and co-workers found clumps of gravitationally bound grains on such maps~\cite{johansen2007}. These clumps are counted and their sizes are measured, as on a photograph. A continuum model, on the other hand, gives fields. To compare the two, the simulations must be measured as images are measured.

To answer these two questions, we combine two tools. The first one solves the equations of the model. It is a second-order adaptive finite-volume scheme, designed for scalar conservation laws with a concave, bell-shaped flux. The volume-filling flux $\rho(1-\rho)$ has exactly this shape. The second tool is an image processing chain. It detects and measures the clumps on each simulated field. Finally, we repeat the simulation for 192 combinations of the model parameters. The theory is thus tested in very different situations.

The answer is positive for the start of clump formation. Linear theory predicts which cases form clumps, at what time, and with what initial spacing. However, it does not describe the slow coarsening that follows.

Section~\ref{sec:modele} presents the model, its linear stability and its numerical solution, and then verifies the code. Section~\ref{sec:protocole} describes the protocol: the initial condition, the image processing, the measured quantities and the simulations performed. In Section~\ref{sec:ref}, a reference case shows the stages of clump formation. Section~\ref{sec:campagne} then varies the strength of gravity and the amount of grains, and tests the predictions of linear theory for each combination. The sensitivity analysis of Section~\ref{sec:sobol} finds which of the two parameters matters most. Finally, Section~\ref{sec:discussion} links these results to two previous discrete studies~\cite{trivino2025,trivino2026}.

\section{Model}\label{sec:modele}

We choose this model for four reasons. First, it is the simplest model that contains the ingredients of clump formation: attraction, spreading and crowding. Second, its solutions stay bounded. The simulations can therefore run for a long time, and the clumps have a finite size that can be measured on an image. Third, its linear stability is explicit, which gives predictions that are easy to test. Fourth, it depends on only two dimensionless parameters, so the whole parameter plane can be explored. The model also has a technical advantage. Its flux $\rho(1-\rho)$ is the flux of road traffic models, for which we know how to build schemes that respect the bounds $0\le\rho\le1$.

This section presents the equations and then their linear stability. It then describes their numerical solution and verifies the code.
\subsection{Dimensionless equations}

We look from above at a piece of the dust layer, which we assume to be thin. The problem is therefore two-dimensional. We write $\rho(t,\mathbf{x})\in[0,1]$ for the volume fraction of grains, scaled by maximum packing. It is the part of space occupied by grains at point $\mathbf{x}$. The value $\rho=0$ is an empty region, and the value $\rho=1$ is a region where the grains are packed as much as possible. We write $\Phi$ for the gravitational potential created by the grains. Each grain slides toward the bottom of the wells of $\Phi$ with the velocity $\mathbf{v}=-\chi\nabla\Phi$, which is set by the gas drag. Lengths are scaled by the range $\ell$ of gravity. Time is scaled by the time $\ell^2/D$ that turbulence, with diffusivity $D$, needs to spread the grains over this range. The model reads
\begin{align}
  \partial_t \rho + \nabla\!\cdot\!\big(\rho(1-\rho)\,\mathbf{v}\big) &= \Delta\rho, \qquad \mathbf{v}=-\chi\nabla\Phi, \label{eq:rho}\\
  \Delta\Phi - \Phi &= \rho. \label{eq:c}
\end{align}
Equation~\eqref{eq:rho} contains two motions. The term $\Delta\rho$ describes the turbulent shaking of the grains, which spreads them out. The term with $\mathbf{v}$ describes their fall toward the potential wells. Equation~\eqref{eq:c} is the equation of gravity. Without the term $-\Phi$, it would be the Poisson equation of Newtonian gravity. This term limits the range of the attraction to about one length unit: this is called screened gravity. It is a simple way to mimic the effect of the rotation of the disk, which opposes collapse at large scales. The domain is the square $\Omega=[0,L]^2$, with homogeneous Neumann boundary conditions. No grain enters or leaves the domain. Gravity adjusts instantly to the position of the grains, so Eq.~\eqref{eq:c} has no time derivative.

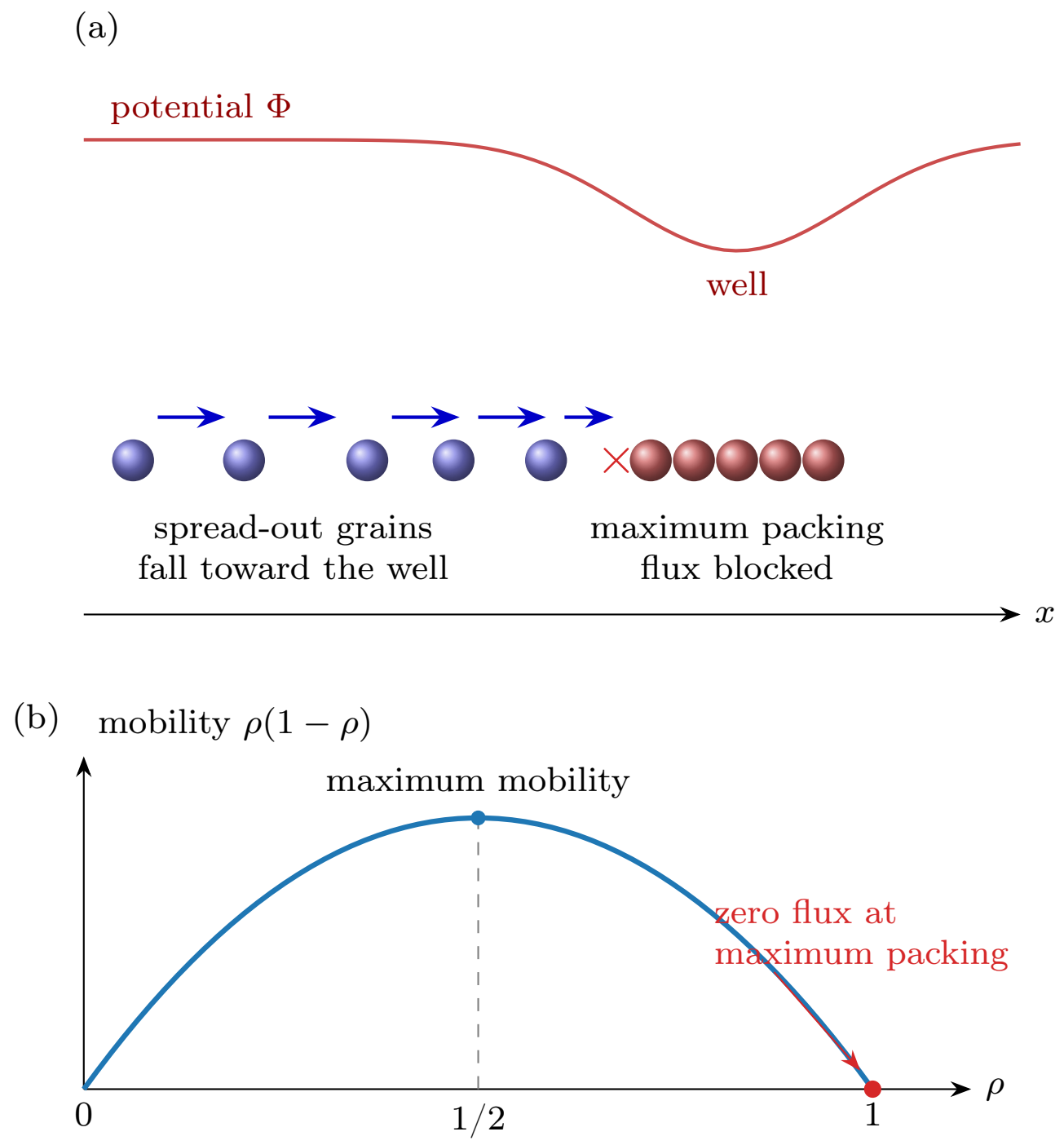
\begin{figure}[htb]
\centering
\begin{tikzpicture}[x=1cm,y=1cm,font=\footnotesize]
  \node[anchor=west] at (-0.2,4.0) {(a)};
  \draw[seisR!70, thick, domain=0:7.6, samples=80] plot (\x,{3.1-0.9*exp(-((\x-5.3)^2)/1.6)});
  \node[seisR!80!black, anchor=west] at (0.1,3.35) {potential $\Phi$};
  \node[seisR!80!black, anchor=north] at (5.3,2.15) {well};
  \foreach \x in {0.4,1.3,2.3,3.0} {\shade[ball color=seisB!50] (\x,0.5) circle (0.17);}
  \foreach \x in {0.4,1.3,2.3,3.0} {\draw[-{Stealth}, seisB, thick] (\x+0.2,0.85) -- (\x+0.75,0.85);}
  \foreach \x in {4.6,4.95,5.3,5.65,6.0} {\shade[ball color=seisR!60] (\x,0.5) circle (0.17);}
  \shade[ball color=seisB!50] (3.75,0.5) circle (0.17);
  \draw[-{Stealth}, seisB, thick] (3.9,0.85) -- (4.3,0.85);
  \node[mplC3, font=\large\bfseries] at (4.32,0.5) {$\times$};
  \node[align=center, anchor=north] at (5.3,0.18) {maximum packing\\ flux blocked};
  \node[align=center, anchor=north] at (1.7,0.18) {spread-out grains\\ fall toward the well};
  \draw[-{Stealth}] (0,-0.75) -- (7.6,-0.75) node[right] {$x$};
  \begin{scope}[yshift=-4.6cm]
    \node[anchor=west] at (-0.7,3.0) {(b)};
    \draw[-{Stealth}] (0,0) -- (7.2,0) node[right] {$\rho$};
    \draw[-{Stealth}] (0,0) -- (0,2.7) node[anchor=south west] {mobility $\rho(1-\rho)$};
    \draw[mplC0, very thick, domain=0:6.4, samples=60] plot (\x,{2.2*4*(\x/6.4)*(1-\x/6.4)});
    \draw[dashed, gray] (3.2,0) -- (3.2,2.2);
    \node[below] at (0,0) {0};
    \node[below] at (3.2,0) {1/2};
    \node[below] at (6.4,0) {1};
    \fill[mplC0] (3.2,2.2) circle (0.06);
    \node[anchor=south, align=center] at (3.2,2.25) {maximum mobility};
    \fill[mplC3] (6.4,0) circle (0.07);
    \node[anchor=west, align=left, mplC3] at (5.0,1.25) {zero flux at\\ maximum packing};
    \draw[-{Stealth}, mplC3] (5.6,0.95) -- (6.3,0.15);
  \end{scope}
\end{tikzpicture}
\caption{The grains fall into the gravity well that they dig, but they cannot pile up there without limit. (a)~The grains slide toward the gravitational potential well. A grain cannot enter a clump that is already at maximum packing. (b)~The mobility $\rho(1-\rho)$ therefore vanishes at $\rho=1$, which bounds the volume fraction in the clumps.}
\label{fig:mecanisme}
\end{figure}

Figure~\ref{fig:mecanisme} shows the role of volume filling. The mobility $\rho(1-\rho)$ is maximum at $\rho=1/2$. It vanishes at maximum packing~\cite{painter2002}. A grain therefore cannot enter a saturated clump, and the volume fraction there stays below 1. This is the mathematical form of crowding.

Two dimensionless numbers control the problem. The first one is the gravitational number $\chi=4\pi G\rho_{\max}t_s\ell^2/D$. Here, $G$ is the gravitational constant, $\rho_{\max}$ is the mass density of packed grains and $t_s$ is the stopping time of a grain in the gas. This number compares the fall of the grains toward the wells with their spreading by turbulence. The larger $\chi$ is, the more gravity wins. The second one is the mean volume fraction $\rhob=|\Omega|^{-1}\int_\Omega\rho\,d\mathbf{x}$. It measures the total amount of grains. It does not change in time, because no grain appears or disappears. The size of the domain is $L=16$ in the whole study, that is 16 times the range of gravity.

This model is exactly the volume-filling Keller--Segel model~\cite{painter2002}, with $c=-\Phi$. The code therefore solves these equations in the form $\partial_t\rho+\nabla\cdot(\chi\rho(1-\rho)\nabla c)=\Delta\rho$ and $-\Delta c+c=\rho$.

\subsection{Linear stability of the base state}

The homogeneous base state $\rho=\rhob$, $\Phi=-\rhob$, is an exact solution of the model. In this state, the grains are spread uniformly and there is no clump. To know whether this state survives, we perturb it with a normal mode $\cos(k_x x)\cos(k_y y)$ of wavenumber $k$. The dispersion relation reads
\begin{equation}
  \sigma(k) = -k^2 + \Lambda\,\frac{k^2}{1+k^2},\qquad \Lambda=\chi\,\rhob(1-\rhob). \label{eq:sigma}
\end{equation}
Diffusion stabilizes short wavelengths, while gravity destabilizes long ones. Stability therefore depends only on the control parameter $\Lambda$. This parameter measures the effective strength of the attraction. It combines the strength of gravity $\chi$ and the mobility of the grains $\rhob(1-\rhob)$. A mode with $\sigma(k)>0$ grows exponentially, and a mode with $\sigma(k)<0$ fades away.

\begin{figure}[htb]
\centering
\begin{tikzpicture}
\begin{axis}[name=ax, width=\columnwidth, height=5.2cm, xmin=0, xmax=2.1, ymin=-1.3, ymax=1.5,
  xlabel={fineness of the ripple: wavenumber $k=2\pi/\lambda$}, ylabel={growth rate $\sigma$},
  axis lines=left, tick label style={font=\scriptsize}, label style={font=\footnotesize},
  xtick={0,0.5,1,1.5,2}, ytick={-1,0,1}]
  \fill[mplC1!12] (axis cs:0,0) rectangle (axis cs:1.789,1.5);
  \addplot[domain=0:2.1, samples=120, mplC0, very thick] {-x^2 + 4.2*x^2/(1+x^2)};
  \addplot[domain=0:1.3, samples=80, mplC2, very thick, dashed] {-x^2 + 0.8*x^2/(1+x^2)};
  \addplot[gray, thin] coordinates {(0,0) (2.1,0)};
  \addplot[mplC3, dashed] coordinates {(1.0247,0) (1.0247,1.101)};
  \addplot[only marks, mark=*, mark size=1.6pt, mplC3] coordinates {(0.25,0.185) (1.0247,1.101) (1.95,-0.477)};
  \node[font=\scriptsize, anchor=north east, mplC0, align=right] at (axis cs:2.08,1.48) {strong gravity, $\Lambda=4.2$:\\ clumps form};
  \node[font=\scriptsize, anchor=west, mplC2, align=left] at (axis cs:0.03,-0.8) {weak gravity, $\Lambda=0.8$:\\ the layer stays uniform};
  \node[font=\scriptsize, anchor=south, mplC3] at (axis cs:1.0247,1.13) {$k^*$};
  \node[font=\scriptsize, anchor=west, mplC1!80!black] at (axis cs:0.42,0.4) {$\sigma>0$: the ripple grows};
  \node[font=\scriptsize, anchor=north west, gray!60!black] at (axis cs:1.0,-0.1) {$\sigma<0$: it fades away};
  \coordinate (pA) at (axis cs:0.25,0.185);
  \coordinate (pB) at (axis cs:1.0247,1.101);
  \coordinate (pC) at (axis cs:1.95,-0.477);
\end{axis}
\pgfmathsetseed{7}
\foreach \col/\lam/\amp/\txt [count=\c] in {
    0/4.6/0.5/{long ripple\\ (slow)},
    1/0.75/0.5/{ripple $\lambda^*$\\ (the fastest)},
    2/0.2/0.15/{fine ripple\\ (fades away)}} {
  \begin{scope}[xshift=\col*2.7cm-0.7cm, yshift=5.25cm]
    \fill[gray!6] (0,0) rectangle (2.3,0.6);
    \foreach \n in {1,...,380} {
      \pgfmathsetmacro{\x}{2.3*rnd}
      \pgfmathsetmacro{\y}{0.04+0.52*rnd}
      \pgfmathtruncatemacro{\keep}{rnd < 0.5+\amp*cos(360*\x/\lam)}
      \ifnum\keep=1 \fill[black!75!orange] (\x,\y) circle (0.022); \fi
    }
    \node[font=\scriptsize, align=center, anchor=south] (lab\c) at (1.15,0.62) {\txt};
    \coordinate (s\c) at (1.15,0);
  \end{scope}}
\draw[-{Stealth}, gray] (s1) -- (pA);
\draw[-{Stealth}, gray] (s2) -- (pB);
\draw[-{Stealth}, gray] (s3) -- (pC);
\end{tikzpicture}
\caption{Which ripples of the grain layer grow? Top: three ripples of the layer, drawn grain by grain. Bottom: their growth rate, given by the dispersion relation~\eqref{eq:sigma}. When gravity is strong enough (solid line), a band of ripples grows. The fastest one, with wavelength $\lambda^*=2\pi/k^*$, should set the spacing of the first clumps. Ripples that are too fine are erased by turbulence. When gravity is weak (dashed line), all ripples fade away.}
\label{fig:dispersion}
\end{figure}
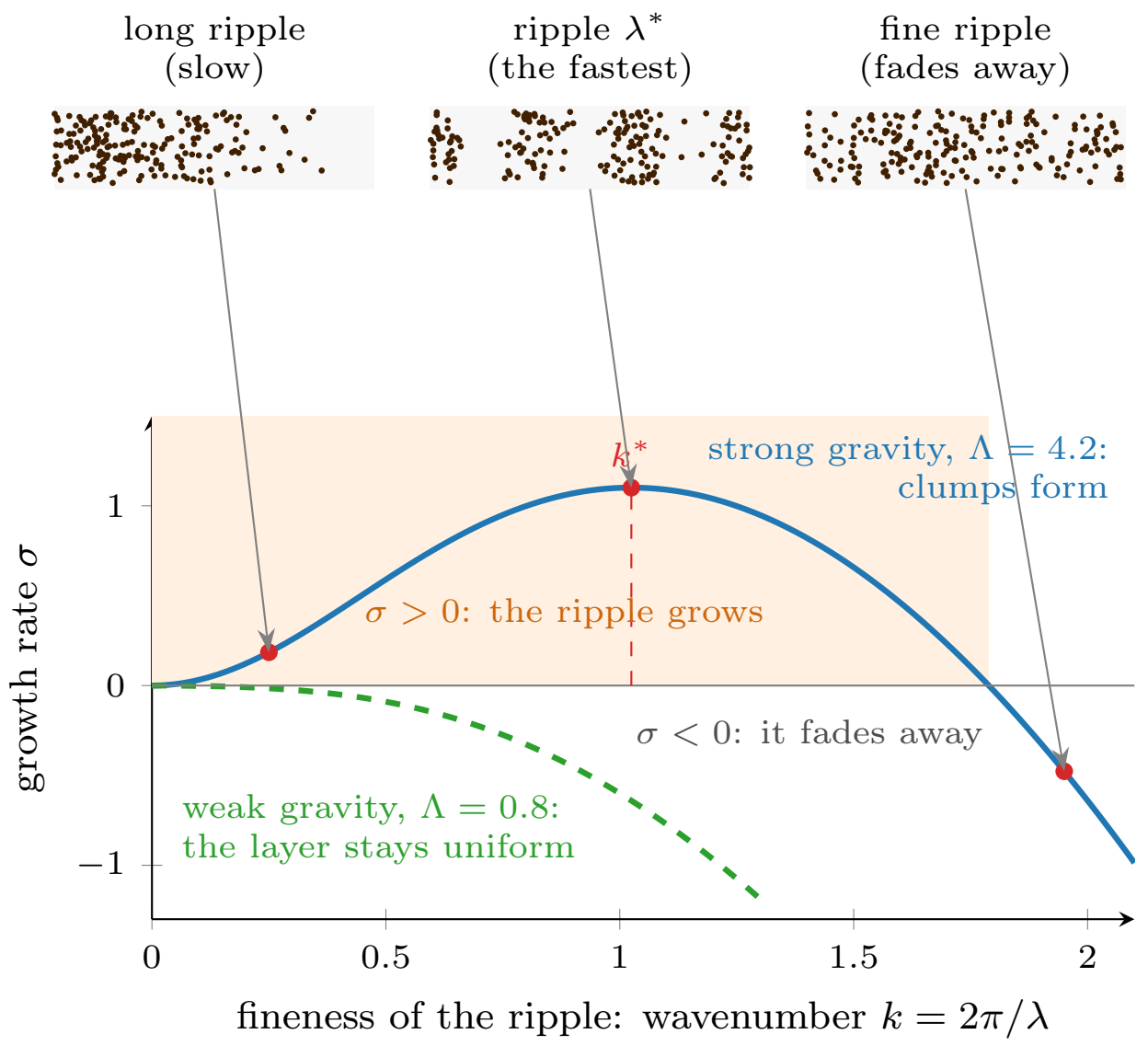

Figure~\ref{fig:dispersion} shows the two possible situations. The Neumann conditions only allow the wavenumbers $k=\pi\sqrt{m^2+n^2}/L$. The base state is therefore unstable if $\Lambda > 1+(\pi/L)^2$. In the $(\chi,\rhob)$ plane, the marginal stability curve reads $\rhob=\tfrac12\big(1\pm\sqrt{1-4(1+\pi^2/L^2)/\chi}\big)$. It exists only for $\chi>4(1+\pi^2/L^2)$. Above the threshold, the most amplified mode has the wavenumber $k^{*}=(\sqrt\Lambda-1)^{1/2}$. Its growth rate is $\sigma^*=(\sqrt\Lambda-1)^2$ and its wavelength is $\lambda^*=2\pi/k^*$. These two quantities give the predictions tested below. The most amplified mode grows faster than all the others. It should therefore set the spacing of the first clumps, of the order of $\lambda^*$. Its growth rate $\sigma^*$ should set the time at which they appear, of the order of a few $1/\sigma^*$. In short, clumps form only if the attraction $\Lambda$ is above a threshold close to 1.

\subsection{Finite-volume discretization}

The discretization must conserve mass and keep the bounds $0\le\rho\le1$. For this, we use finite volumes on a non-uniform Cartesian mesh, with mesh sizes $h^x_i$ and $h^y_j$. The domain is divided into small rectangles, called cells. The unknowns are the averages $\rho_{ij}$ over each cell. The computation follows the exchanges of grains between neighboring cells. What leaves a cell enters its neighbor, so the total amount of grains is exactly conserved.

\begin{figure}[htb]
\centering
\begin{tikzpicture}[font=\footnotesize]
  \node[anchor=west] at (-0.3,3.0) {(a)};
  \pgfmathsetseed{3}
  \foreach \a/\b/\m in {0/1.4/0.40,1.4/2.4/0.62,2.4/3.1/1.10,3.1/3.8/1.70,3.8/4.8/1.92,4.8/6.2/1.95,6.2/7.6/1.95} {
    \pgfmathtruncatemacro{\ng}{round(28*\m*(\b-\a))}
    \foreach \n in {1,...,\ng} {\fill[black!70!orange] ({\a+0.04+(\b-\a-0.08)*rnd},{0.04+(\m-0.08)*rnd}) circle (0.022);}
    \draw[gray, dashed, thick] (\a,\m) -- (\b,\m);
  }
  \foreach \x in {0,1.4,2.4,3.1,3.8,4.8,6.2,7.6} {\draw[gray] (\x,0) -- (\x,2.3);}
  \draw[thick] (0,0) -- (7.6,0);
  \draw[mplC0, very thick] (0,0.35) -- (1.4,0.45);
  \draw[mplC0, very thick] (1.4,0.5) -- (2.4,0.75);
  \draw[mplC0, very thick] (2.4,0.85) -- (3.1,1.35);
  \draw[mplC0, very thick] (3.1,1.55) -- (3.8,1.85);
  \draw[mplC0, very thick] (3.8,1.9) -- (4.8,1.95);
  \draw[mplC0, very thick] (4.8,1.95) -- (6.2,1.95);
  \draw[mplC0, very thick] (6.2,1.95) -- (7.6,1.95);
  \fill[mplC3] (3.1,1.35) circle (0.05);
  \fill[mplC3] (3.1,1.55) circle (0.05);
  \node[mplC3, font=\scriptsize] at (3.3,1.2) {$a$};
  \node[mplC3, font=\scriptsize] at (2.92,1.68) {$b$};
  \draw[-{Stealth}, mplC3, thick] (2.75,2.5) -- (3.45,2.5) node[right, font=\scriptsize] {grain flux $G(a,b)$};
  \draw[mplC3, dotted] (3.1,1.6) -- (3.1,2.45);
  \draw[decorate, decoration={brace, mirror, amplitude=3pt}] (2.4,-0.05) -- (3.1,-0.05) node[midway, below=2pt] {cell};
  \node[anchor=north west, align=left, font=\scriptsize] at (0.05,2.25) {\textcolor{gray}{dashed: average}\\ \textcolor{mplC0}{solid: line}};
  \node[anchor=north, font=\scriptsize] at (6.2,-0.1) {dense region (clump)};
  \node[anchor=north, font=\scriptsize] at (0.9,-0.1) {dilute region};
  \begin{scope}[yshift=-4.6cm]
    \node[anchor=west] at (-0.3,3.7) {(b)};
    \begin{scope}[xshift=0.6cm]
      \foreach \x in {0,0.55,0.62,0.7,0.8,0.9,1.0,1.1,1.2,1.3,1.38,1.45,2.0,2.05,2.15,2.25,2.35,2.45,2.55,2.65,2.75,2.85,2.95,3.05,3.15,3.25,3.3,3.5} {\draw[gray!60, very thin] (\x,0) -- (\x,3.5);}
      \foreach \y in {0,0.3,0.4,0.5,0.6,0.7,0.8,0.9,1.0,1.1,1.2,1.3,1.4,1.5,2.1,2.2,2.3,2.4,2.5,2.6,2.7,2.8,2.9,3.0,3.1,3.2,3.3,3.5} {\draw[gray!60, very thin] (0,\y) -- (3.5,\y);}
      \pgfmathsetseed{11}
      \foreach \cx/\cy in {1.0/1.0,2.4/2.5,0.9/2.9,2.9/0.7} {
        \foreach \n in {1,...,70} {\fill[black!70!orange] ({\cx+0.36*sqrt(rnd)*cos(360*rnd)},{\cy+0.36*sqrt(rnd)*sin(360*rnd)}) circle (0.022);}
      }
      \draw (0,0) rectangle (3.5,3.5);
      \draw[-{Stealth}] (4.0,3.3) node[right, font=\scriptsize] {clump of grains} -- (2.75,2.65);
      \draw[-{Stealth}] (4.0,1.75) node[right, font=\scriptsize] {wide cells} -- (1.75,1.8);
    \end{scope}
    \node[anchor=west, align=left, text width=3.3cm, font=\scriptsize] at (4.6,0.15) {Each clump refines whole lines of the mesh. With many clumps, fine cells are needed almost everywhere.};
  \end{scope}
\end{tikzpicture}
\caption{Discretization. (a)~Each cell contains an amount of grains, known only through its average (dashed lines). In each cell, a straight line (solid line) reconstructs the profile. The grain flux $G(a,b)$ through an interface is computed from the two values read on these lines, $a$ on the left and $b$ on the right. (b)~Adapted mesh around four clumps of grains. The fine cells spread along the lines of the mesh.}
\label{fig:discretisation}
\end{figure}
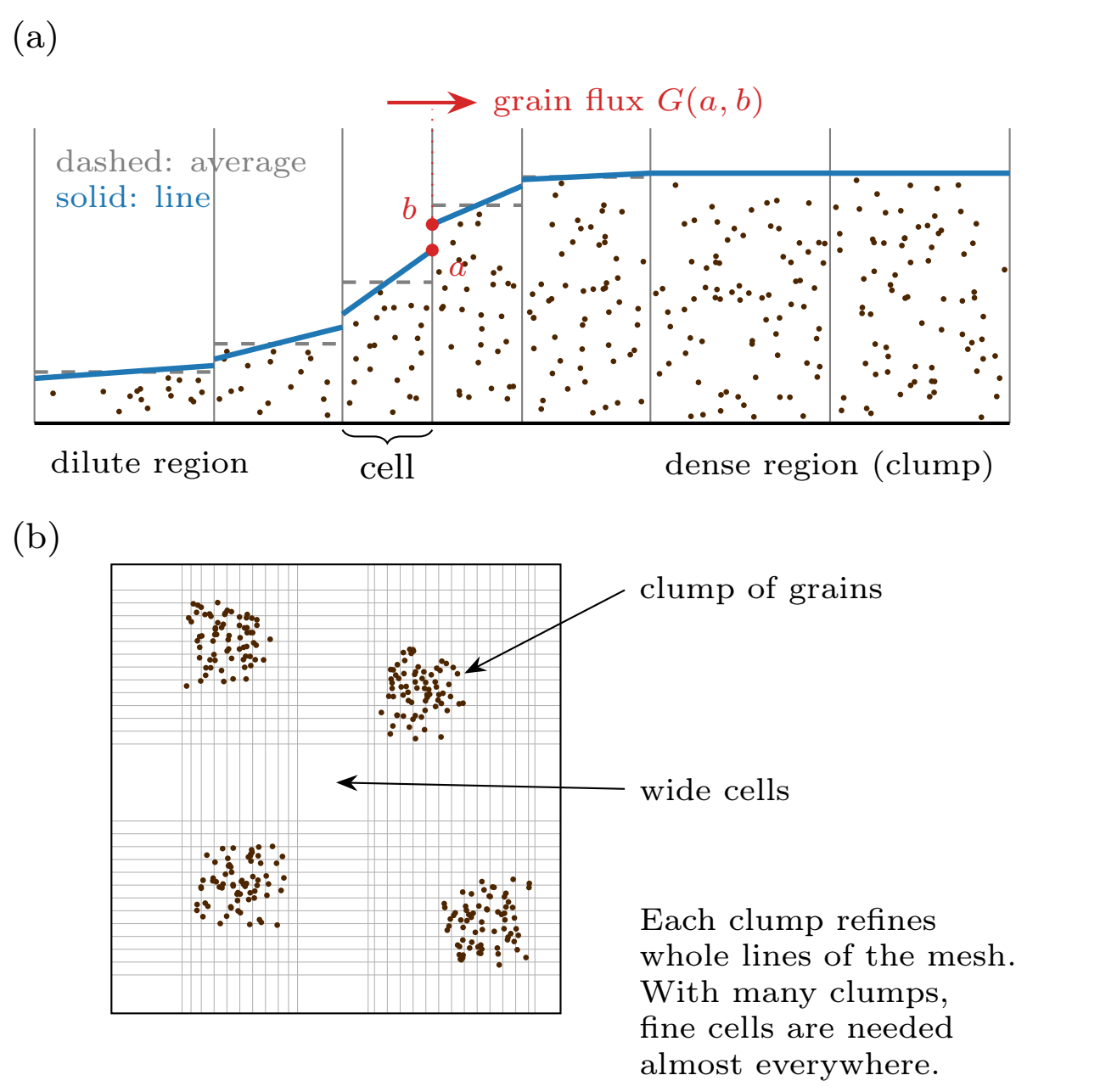

To compute what crosses an interface, we need to know $\rho$ on both sides of it. But we only know the average of $\rho$ in each cell. The simplest solution takes this average as the value at the edge. The profile of $\rho$ then becomes a staircase, and the steep fronts of the clumps are artificially smeared. Instead, we reconstruct in each cell a straight line that passes through the average (Fig.~\ref{fig:discretisation}a). Its slope is estimated from the two neighboring cells. We then read the value at the edge of the cell on this line. An interface thus separates two values, $a$ on the left and $b$ on the right, read on the lines of the two cells. This method, called MUSCL reconstruction~\cite{vanleer1979}, makes the scheme second-order accurate.

However, a badly chosen slope could create new values, for example $\rho>1$ at the top of a clump. To avoid this, we compare the slope toward the left cell and the slope toward the right cell. We keep the smaller of the two. If they have opposite signs, the cell is a peak or a valley, and we take a zero slope. This choice, called the minmod limiter, creates no new maximum or minimum.

We still have to decide which flux crosses the interface between the states $a$ and $b$. We use the Godunov flux. It solves exactly the local problem of two states that meet, for the drift flux $g(\rho)=u\,\rho(1-\rho)$. The fall velocity $u=-\chi\,\partial\Phi$ is evaluated at the interface with a centered difference. The function $g$ is concave if $u>0$ and convex if $u<0$. The numerical flux is therefore the minimum of $g$ on $[a,b]$ if $a\le b$, and its maximum on $[b,a]$ otherwise. The diffusive flux is centered, and the total flux is zero on the boundary.

The potential is obtained by discretizing Eq.~\eqref{eq:c} with finite volumes on the same mesh. Its matrix depends only on the mesh. A sparse LU decomposition, computed once per remeshing, is therefore enough for the whole time between two remeshings.

Time integration uses Heun's second-order SSP Runge--Kutta scheme~\cite{gottlieb2001}. Each time step combines two explicit Euler steps. A single Euler step respects the bounds $0\le\rho\le1$, and the combination keeps this property. This is the meaning of SSP (strong stability preserving). The time step must stay small enough so that matter does not cross more than one cell per step. This constraint, called the CFL condition, reads
\begin{equation}
  \Delta t = \frac{0.9}{2}\,\min_{ij}\Big[\sum_{d\in\{x,y\}}\Big(\frac{|u_d|}{h^d}+\frac{2}{(h^d)^2}\Big)\Big]^{-1}.
\end{equation}
The factor $1/2$ prevents the reconstruction from creating oscillations. The scheme is then conservative and keeps the bounds. In developed clumps, the diffusive term $2/h^2$ sets the time step.

\subsection{Mesh adaptation}

The interfaces of the clumps are steep and they move. A uniform mesh that is fine everywhere would be expensive. An adaptive mesh uses fine cells where the solution changes quickly, and coarse cells elsewhere. We therefore adapt the mesh every $T_w=0.5$ time units, direction by direction. The desired cell size is given by a metric. It asks for small cells where $\rho$ is strongly curved, that is at the interfaces of the clumps. This metric, called the Hessian metric, reads
\begin{equation}
  M = \min\!\big(\max(|\partial_{dd}\rho|/\varepsilon,\,h_{\max}^{-2}),\,h_{\min}^{-2}\big).
\end{equation}
Its maximum in the other direction gives a 1D metric for each direction. This metric is dilated over the distance $\delta = c_\delta(v_{\max}T_w+\sqrt{2T_w})$, which covers advection and diffusion during $T_w$. The nodes then follow the equidistribution principle for the metric length $\int\sqrt{M}\,dx$. Finally, a gradation criterion limits the size ratio of two neighboring cells to 1.5.

The solution is transferred to the new mesh by a second-order conservative projection. Each new average is the exact integral of the MUSCL reconstruction of the old solution. Mass is thus conserved to machine precision, and the bounds are kept. The settings are $\varepsilon=\num{2e-3}$, $h_{\min}=0.15$, $h_{\max}=0.5$ and $c_\delta=1$, with at most 160 cells per direction. Figure~\ref{fig:discretisation}b shows the limit of this approach. On a Cartesian mesh, each clump refines whole lines.

\subsection{Verification}\label{sec:verif}

Before we interpret the simulations, we must make sure that the code solves the model correctly. Otherwise, a programming error could look like a physical effect. A test solution checks the order of convergence. The dispersion relation checks the coupling between the two equations. Finally, the mass balance checks conservation.

\begin{figure}[htb]
  \includegraphics[width=\columnwidth]{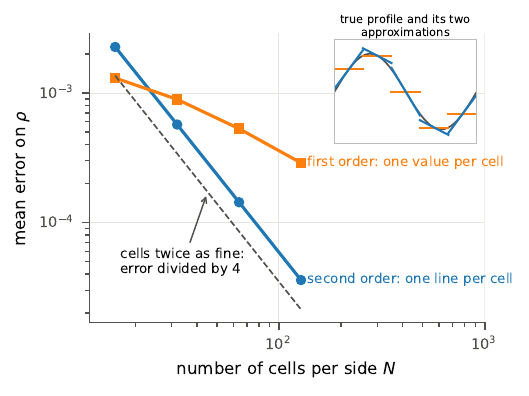}
  \caption{The code converges at second order. Mean error on $\rho$ at $t=0.5$, compared with the test solution, as a function of the number of cells per side. With one straight line per cell (second order), the error is divided by 4 when the cells are twice as fine. With one single value per cell (first order), it decreases much more slowly. The inset shows the two approximations of the same profile.}
  \label{fig:conv}
\end{figure}

A test solution is a solution chosen in advance, so it is known exactly. We add to the equations the source term that makes it exact, and then we compare the computation with this known solution. Here, the test solution is $\rho_e=\tfrac12+b(t)\cos(kx)\cos(ky)$, with $k=\pi/L$, $L=4$ and $b(t)=0.25+0.1\sin t$. This field satisfies the Neumann conditions. The corresponding exact potential is $\Phi_e=-\tfrac12-b\cos(kx)\cos(ky)/(1+2k^2)$, and the source term is computed analytically. Figure~\ref{fig:conv} shows an error that scales as $N^{-2}$. The error is therefore divided by 4 when the cell size is divided by 2. The observed order is 2.00 between 64 and 128 cells. The first-order scheme only reaches order 0.88.

The order of convergence says nothing about the coupling between the fall of the grains and gravity. We therefore perturb the state $\rhob=0.3$ with the single mode $(m,n)=(3,2)$, with amplitude \num{e-5}, $\chi=10$ and $L=16$. The simulated growth rate is 0.1999. It agrees with Eq.~\eqref{eq:sigma} to within \num{1.1e-4}.

Conservation remains. The relative error on mass never exceeds \num{4.4e-15} over the 192 simulations of the campaign, remeshings included. The volume fraction stays between \num{1.6e-7} and $1-\num{2e-7}$. The balances in the next sections are therefore reliable.

\subsection{Mesh adaptation pays off little when the clumps fill the domain}\label{sec:adapt}

The adaptive mesh is useful only if it reduces the cost for the same accuracy. We therefore compare it with uniform meshes on a typical case. The result also sets the settings used for the campaigns.

\begin{figure}[htb]
\centering
\includegraphics[width=\columnwidth]{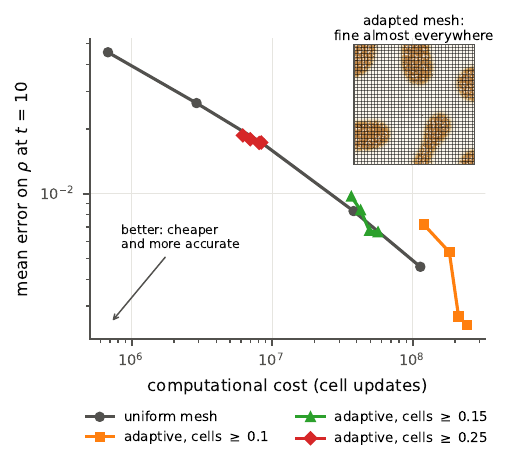}
\caption{The adaptive mesh does barely better than the uniform mesh. Mean error on $\rho$ at $t=10$ as a function of the computational cost, for $\chi=20$ and $\rhob=0.3$. A good mesh is at the bottom left. Each adaptive curve fixes the smallest allowed cell $h_{\min}$ and varies the tolerance $\varepsilon$ from \num{5e-4} to \num{5e-2}. The reference is a uniform mesh of $256^2$ cells. The inset shows why: the clumps are everywhere, and the adapted mesh is fine almost everywhere.}
\label{fig:adapt}
\end{figure}

The test case is $\chi=20$, $\rhob=0.3$, simulated up to $t=10$. The computational cost is the total number of cell updates. Figure~\ref{fig:adapt} shows that the minimum cell size $h_{\min}$ sets the error, while the tolerance $\varepsilon$ has little effect. For $h_{\min}=0.25$, for example, the error stays between \num{1.7e-2} and \num{1.9e-2} over two decades of $\varepsilon$.

The geometry of the pattern explains this result. Eleven clumps occupy the whole domain. Almost every line of the mesh crosses an interface, as shown in Fig.~\ref{fig:discretisation}b. The 1D metric therefore imposes $h_{\min}$ almost everywhere. The best adaptive mesh reaches an error of \num{2.4e-3} for a cost of \num{2.5e8}. For the same cost, a uniform mesh would reach \num{3.0e-3}. The gain is only about \SI{20}{\percent}. In other words, the clumps are so numerous that the mesh must be fine almost everywhere. An unstructured mesh would follow these closed and scattered interfaces better.

The campaigns finally use $h_{\min}=0.15$ and $\varepsilon=\num{2e-3}$. The error is then \num{6.7e-3}, that is \SI{2}{\percent} of the mean volume fraction. The number of clumps is exact.

\section{Protocol}\label{sec:protocole}

\begin{figure*}[htb]
\centering
\begin{tikzpicture}[font=\footnotesize,
  boite/.style={draw, rounded corners=2pt, fill=gray!8, align=center, minimum height=0.9cm, inner sep=3pt},
  fl/.style={-{Stealth}, thick}]
  \node[boite, fill=mplC0!12] (init) {almost uniform\\ layer of grains};
  \node[boite, right=0.6cm of init, fill=mplC0!12] (signal) {gravity created\\ by the grains $\Phi$};
  \node[boite, right=0.6cm of signal, fill=mplC0!12] (flux) {grain flux\\ between cells};
  \node[boite, right=0.6cm of flux, fill=mplC0!12] (rk) {one step\\ forward in time};
  \node[boite, right=1.2cm of rk, fill=mplC1!15] (adapt) {mesh adaptation\\ (every 0.5 time unit):\\ fine cells at the edges\\ of the clumps};
  \draw[fl] (init) -- (signal);
  \draw[fl] (signal) -- (flux);
  \draw[fl] (flux) -- (rk);
  \draw[fl] (rk.north) -- ++(0,0.45) -| node[pos=0.25, above] {next step} ([xshift=3mm]signal.north);
  \draw[fl] (rk) -- (adapt);
  \draw[fl] (adapt.north) -- ++(0,0.95) -| node[pos=0.25, above] {new mesh} ([xshift=-3mm]signal.north);
  \node[below=1.6cm of signal.south west, anchor=north west, inner sep=0] (i1) {\includegraphics[width=2.2cm]{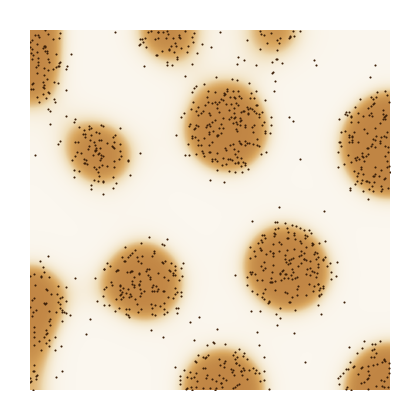}};
  \node[right=0.75cm of i1, inner sep=0] (i2) {\includegraphics[width=2.2cm]{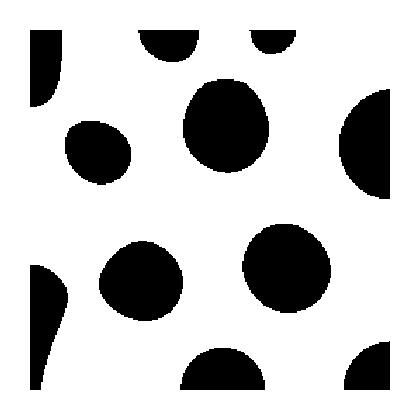}};
  \node[right=0.75cm of i2, inner sep=0] (i3) {\includegraphics[width=2.2cm]{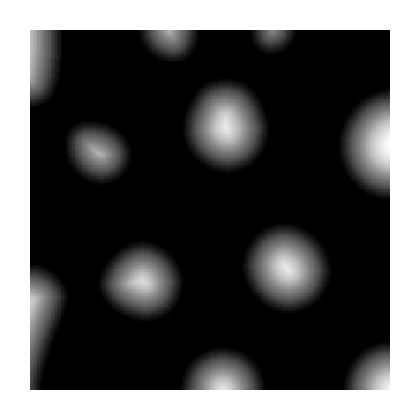}};
  \node[right=0.75cm of i3, inner sep=0] (i4) {\includegraphics[width=2.2cm]{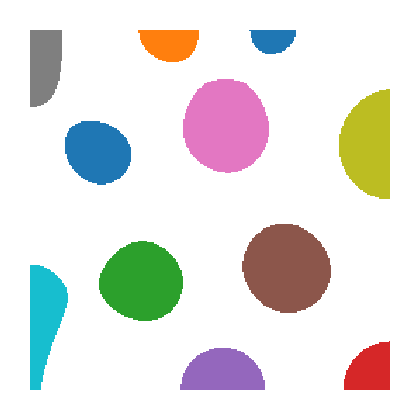}};
  \node[boite, right=0.75cm of i4, fill=mplC2!15] (obs) {measurements:\\ number and area\\ of clumps, $\ton$,\\ spacing $\lambda_{\mathrm{pic}}$};
  \node[below=1pt of i1, align=center] {photograph\\ of the grains};
  \node[below=1pt of i2, align=center] {clumps / background\\ (Otsu threshold)};
  \node[below=1pt of i3, align=center] {distance to the\\ clump edge};
  \node[below=1pt of i4, align=center] {separated clumps\\ (watershed)};
  \draw[fl] (i1) -- (i2);
  \draw[fl] (i2) -- (i3);
  \draw[fl] (i3) -- (i4);
  \draw[fl] (i4) -- (obs);
  \draw[fl, dashed] (rk.south) -- ++(0,-0.7) coordinate (c1) -- (c1 -| i1.north) node[midway, below] {every 2 time units: we take a photograph of the layer} -- (i1.north);
\end{tikzpicture}
\caption{Protocol. Top: the simulation. We compute the gravity created by the grains, then the grain fluxes between cells, and we move one step forward in time. The mesh is adapted regularly. Bottom: the measurement. Every two time units, we take a photograph of the layer, we separate the clumps from the background, and then we separate the clumps that touch each other. The four images are the real steps of the processing, for the reference case at $t=10$.}
\label{fig:boucle}
\end{figure*}
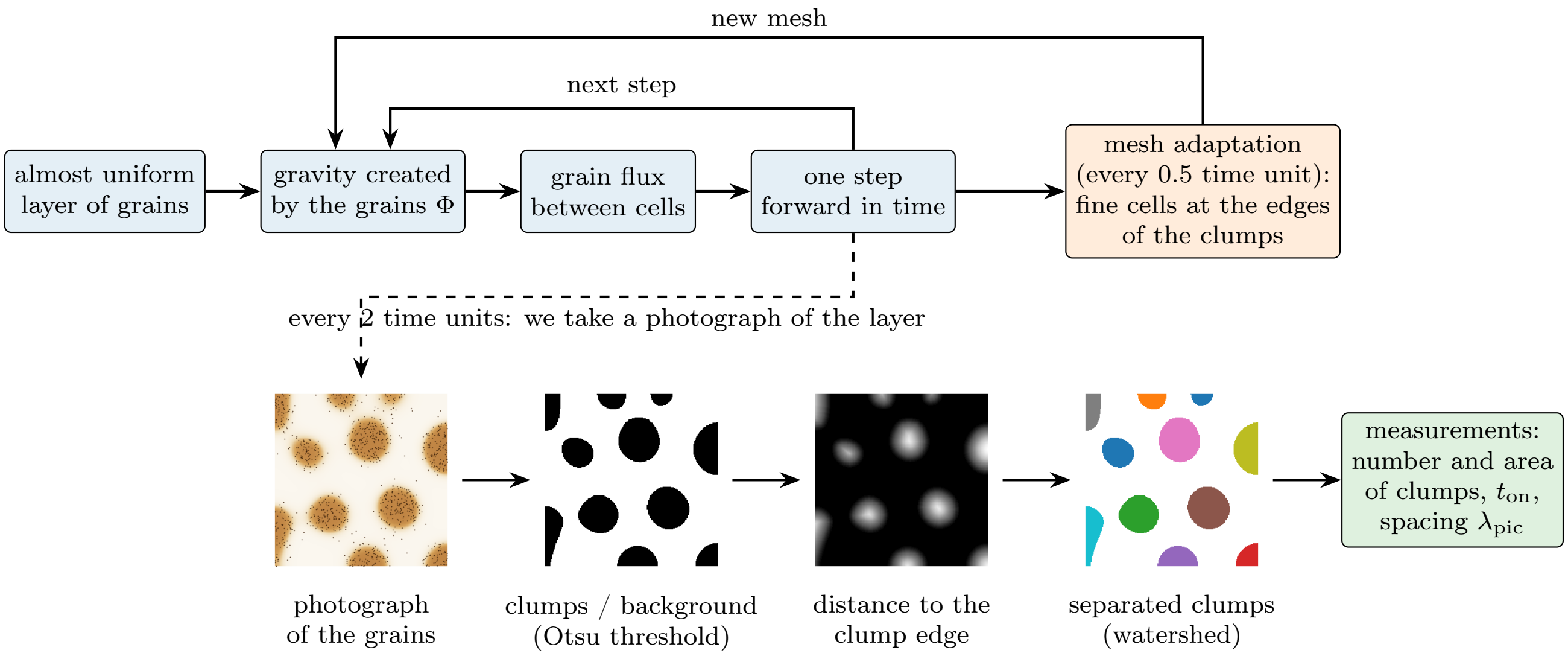

The protocol mimics a repeated observation. We prepare an almost uniform layer of grains. We let it evolve, and we take photographs of it at regular intervals. On each photograph, we find the clumps and we measure them. Finally, we repeat the experiment for many values of the parameters. Each measured quantity answers one of the questions of the introduction. The onset time $\ton$ tells when the clumps appear. The wavelength $\lambda_{\mathrm{pic}}$ tells how far apart they are. The number of clumps over time describes coarsening. Figure~\ref{fig:boucle} summarizes this protocol.

\subsection{Initial condition}

The simulations start from the slightly perturbed base state. This small perturbation mimics the natural irregularities of a real dust layer. The initial field is $\rho_0=\rhob\,(1+0.02\,\xi)$. The perturbation $\xi$ is a sum of normal modes with wavenumber below 2, with Gaussian amplitudes drawn from a random seed. It is normalized to a unit standard deviation. It therefore satisfies the boundary conditions and does not depend on the mesh. The mass is then adjusted to exactly $\rhob L^2$. The seed makes it possible to repeat a simulation with another perturbation, to measure the effect of chance.

\subsection{Image processing}\label{sec:image}

The measured quantities must be those of an image. The goal is to measure the clumps as one would on a photograph or on a density map. Every two time units, the field is therefore projected in a conservative way onto a uniform grid of $256\times256$ pixels.

The image follows a standard processing chain for counting objects on a photograph. It is first smoothed by a Gaussian filter with a standard deviation of 1~pixel, and then binarized by an Otsu threshold~\cite{otsu1979}. This threshold automatically chooses the level that separates the pixels into two classes: the background and the clumps. No manual setting is therefore needed, so the 192 simulations can be processed in the same way. A size filter then removes objects smaller than 12~pixels, and holes are filled. Two clumps that touch then form a single binary object. A watershed transform separates them~\cite{vincent1991}. It cuts the object where it becomes narrow. It works on the distance map, with one seed per regional maximum deeper than 2~pixels. Finally, each clump is described by its area, its mass, its mean volume fraction and its centroid. We also measure its elongation $a/b$, the ratio of its two principal axes. A gamma distribution is fitted to the normalized areas, to describe their spread.

\subsection{Measured quantities}

At the very beginning, the field is almost uniform, and the Otsu threshold would cut the noise. We therefore first measure the peak-to-peak amplitude $\max\rho-\min\rho$. As long as it stays below 0.5, the image is not segmented. The time at which this amplitude reaches 0.5 defines the onset time $\ton$. It is the time at which the clumps become clearly visible.

The wavelength of the pattern is taken from the power spectrum of the image. This spectrum is computed on the basis of the normal modes $\cos(m\pi x/L)\cos(n\pi y/L)$ with a discrete cosine transform. These modes are exactly those of linear theory. The comparison with the dispersion relation is therefore direct. We keep the wavelength of the most energetic mode, $\lambda_{\mathrm{pic}}$, and the mean wavelength $2\pi/\langle k\rangle$. Both are measured on the first segmented image. After $\ton$, a law $n(t)=A\,t^{-\alpha}$ is fitted to the number of clumps with robust least squares. The exponent $\alpha$ characterizes coarsening. The larger $\alpha$ is, the faster the number of clumps decreases.

\subsection{Designs of experiments}

A design of experiments is the list of simulated pairs $(\chi,\rhob)$. The two parameters are explored with two complementary designs. The first one is a grid of 8 values of $\chi$ in geometric progression and 7 values of $\rhob$. Each point is simulated with two seeds, which gives 112 simulations. This grid is used for the maps. The second design is a Latin hypercube of 80 points in the $(\ln\chi,\rhob)$ plane~\cite{mckay1979}. Its points are drawn at random, but they are spread regularly along each axis. It is used for the metamodel. Each simulation stops at $t=40$. A stopping criterion on the residual $\|\partial_t\rho\|_{L^1}/M_0<\num{e-7}$ is included, but it was never reached. The 192 simulations all finished, for a total cost of \SI{2.5}{h} of computation on one core.

\section{Reference case}\label{sec:ref}

A typical case helps us recognize the stages of clump formation before we explore the parameters. It is the common thread of what follows. We choose $\chi=20$ and $\rhob=0.3$, which gives $\Lambda=4.2$. Theory predicts $\sigma^*=1.10$ and $\lambda^*=6.13$.

\begin{figure*}[htb]
  \includegraphics[width=\textwidth]{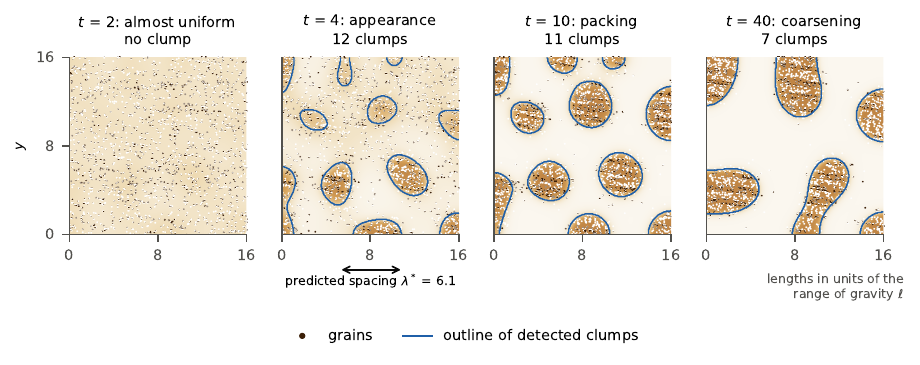}
  \caption{Clump formation in the reference case ($\chi=20$, $\rhob=0.3$). Each dot represents grains, and the shade of the background follows the volume fraction $\rho$. The blue contours outline the clumps detected by the image processing. Clumps appear at $t=4$, with the predicted spacing $\lambda^*$. They become denser, and then some of them merge.}
  \label{fig:snap}
\end{figure*}

\begin{figure*}[htb]
\centering
\includegraphics[width=\textwidth]{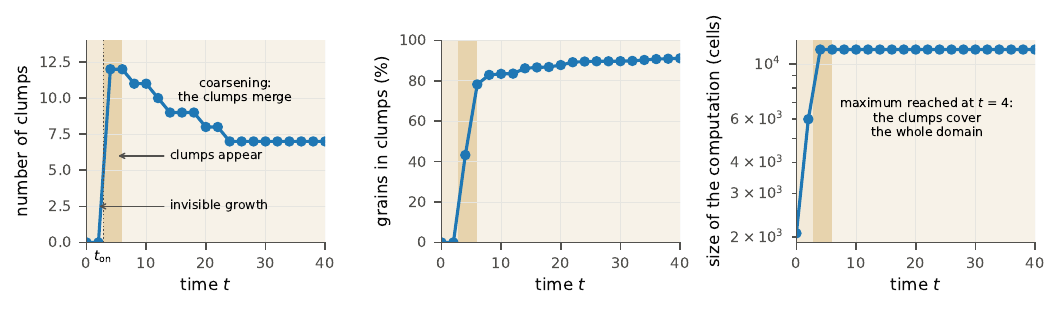}
  \includegraphics[width=\textwidth]{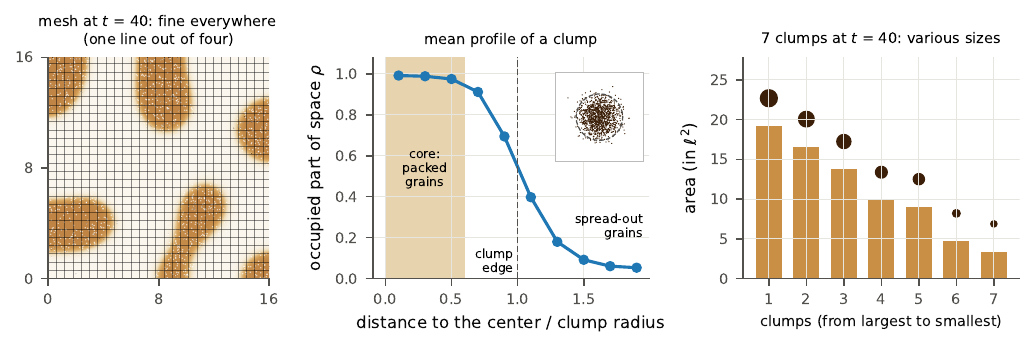}
  \caption{Evolution and final state of the reference case. Top: the number of clumps, the share of grains in the clumps and the size of the computation over time. The colored bands mark the three stages: invisible growth until $\ton$, appearance of the clumps, and then coarsening. Bottom: the state at $t=40$. Left: the mesh, which is fine everywhere. Center: the mean profile of a clump, with a core of packed grains ($\rho\simeq1$), a sharp edge, and then spread-out grains. The distance is scaled by the equivalent radius $r_{\mathrm{eq}}=\sqrt{A/\pi}$ of the clump. Right: the areas of the seven clumps. The disks above the bars have a surface proportional to the area of the clump.}
  \label{fig:refseries}
\end{figure*}

Figures~\ref{fig:snap} and~\ref{fig:refseries} show three regimes. The first one is the linear growth of the perturbation. The initial irregularities grow, but they remain invisible to the eye. The instability starts at $\ton=2.83$, that is $3.1/\sigma^*$. Then comes nonlinear saturation. The densest regions fill up to maximum crowding. Twelve clumps appear almost at the same time at $t=4$, and the most energetic mode then has a wavelength of 6.4, close to $\lambda^*$. The third regime is coarsening. Some clumps merge, and others empty in favor of their neighbors. Eleven clumps remain at $t=10$ and seven at $t=40$, which corresponds to a coarsening exponent of 0.30.

The clumps themselves are almost saturated plateaus. The volume fraction is 0.99 at their center and drops to 0.5 near $r=r_{\mathrm{eq}}$. The interface has a thickness of about $0.5\,r_{\mathrm{eq}}$. At $t=40$, the clumps contain \SI{91}{\percent} of the mass, with a mean volume fraction of 0.91. Their areas vary by a factor of 6 and follow a gamma distribution with shape parameter 3.3. Their mean elongation is 1.5, a sign of ongoing mergers. The mesh reaches its maximum number of cells as early as $t=4$, which confirms the analysis of Section~\ref{sec:adapt}.

\section{Effect of the strength of gravity and of the amount of grains}\label{sec:campagne}

\begin{figure*}[htb]
\centering
\includegraphics[width=\textwidth]{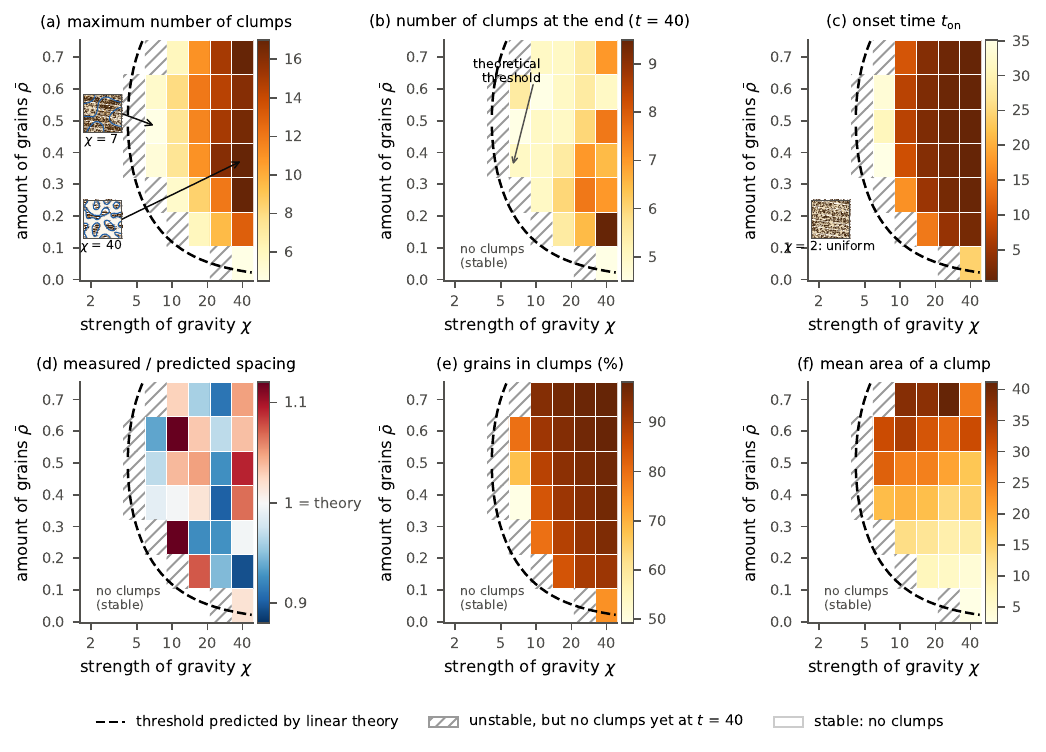}
\caption{Linear theory separates the cases with and without clumps. Six measured quantities in the plane of the two parameters: the strength of gravity $\chi$ and the amount of grains $\rhob$. Each cell is an average over two simulations. The black dashed line is the predicted threshold, $\Lambda=1+(\pi/L)^2$. To its left, no clump forms. The hatched cells are unstable, but too slow to form clumps before $t=40$. The insets show the grains for three cases: few clumps ($\chi=7$), many clumps ($\chi=40$) and a layer that stayed uniform ($\chi=2$). In (d), white corresponds to a spacing equal to the prediction.}
\label{fig:cartes}
\end{figure*}

The reference case suggests that linear theory controls the start of clump formation. A single case is not enough to prove it. We therefore repeat the simulation for many pairs $(\chi,\rhob)$, that is for a stronger or weaker gravity and for more or fewer grains. We must check that this finding holds in the whole $(\chi,\rhob)$ plane. The campaign first compares the simulations with the marginal stability curve. It then tests the predictions on the onset time and on the wavelength. Finally, it looks at what happens to the clumps once they have formed. Figure~\ref{fig:cartes} shows the maps of the grid, and Table~\ref{tab:stats} summarizes the two designs of experiments.

\begin{table}[htb]
  \caption{Summary of the simulations with clumps. The values are medians. The intervals give the 10th and 90th percentiles, or the quartiles for $\alpha$.}
  \label{tab:stats}
  \footnotesize
  \setlength{\tabcolsep}{3pt}
  \begin{ruledtabular}
  \begin{tabular}{lcc}
    & grid & hypercube \\
    \midrule
    unstable cases & 68 & 49 \\
    cases with clumps & 53 & 40 \\
    $\ton\,\sigma^*$ & 3.04 & 3.04 \\
    $\lambda_{\mathrm{pic}}/\lambda^*$ & 0.99 [0.90; 1.11] & 0.98 [0.89; 1.08] \\
    $2\pi\langle k\rangle^{-1}/\lambda^*$ & 0.88 [0.74; 1.00] & 0.87 [0.77; 0.98] \\
    clump area $/(\rhob L^2)$ & 1.00 [0.96; 1.01] & 1.00 [0.97; 1.03] \\
    grains in clumps & 0.93 & 0.92 \\
    $n_{\max}/n(40)$ & 1.8 & 2.0 \\
    exponent $\alpha$ & 0.28 [0.17; 0.35] & 0.36 [0.23; 0.46] \\
    gamma shape of areas & 3.1 & 3.4 \\
  \end{tabular}
  \end{ruledtabular}
\end{table}

\subsection{The marginal stability curve separates the regimes}

Theory first predicts which cases form clumps. No linearly stable case forms clumps. The grid has 68 unstable cases, and 53 of them form clumps before $t=40$. The other 15 cases have a growth rate $\sigma^*\le0.091$. For them, the law $\ton\simeq3/\sigma^*$ found below predicts an onset after $t\simeq33$. These cases are therefore still in linear growth at the end of the simulation; they are not stable. Only one borderline case depends on the seed. The Latin hypercube confirms this result, with 9 cases in the same situation among its 49 unstable cases.

\subsection{Linear theory predicts the onset time}

\begin{figure}[htb]
\centering
\includegraphics[width=\columnwidth]{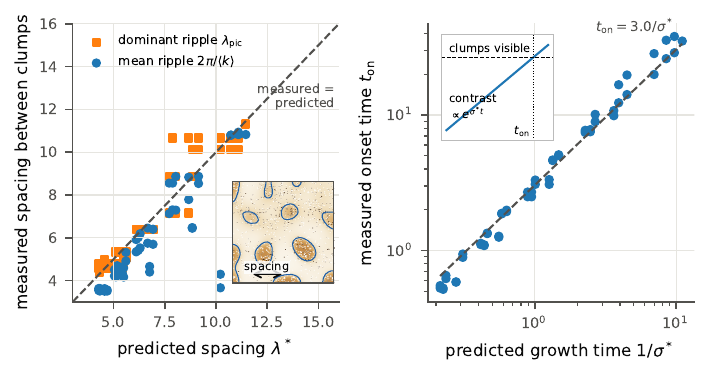}
\caption{Linear theory predicts the spacing and the onset time of the clumps. Each dot is one simulation of the grid. Left: the measured spacing between clumps, on the first image where they appear, as a function of the predicted spacing $\lambda^*$. The inset shows this spacing on the grains of the reference case. Right: the measured onset time as a function of the predicted growth time $1/\sigma^*$. The inset recalls the definition of $\ton$: the contrast grows as $e^{\sigma^*t}$ until the clumps are visible.}
\label{fig:lin}
\end{figure}

Since the threshold is respected, we can test the dynamics itself. The question here is when the clumps appear. Figure~\ref{fig:lin} compares $\ton$ with $1/\sigma^*$ over more than one decade. The correlation coefficient of the logarithms is 0.995. The median of $\ton\,\sigma^*$ is 3.04, with quartiles of 2.64 and 3.27. The two designs of experiments give the same median.

In other words, the clumps appear about three growth times $1/\sigma^*$ after the start. Exponential growth explains this constant. The initial amplitude is about $0.1\,\rhob$ and the threshold is 0.5. The time needed is therefore $\ln(5/\rhob)/\sigma^*$, that is $2.8/\sigma^*$ for $\rhob=0.3$. The measured constant does decrease with $\rhob$. It is 3.3 for $\rhob<0.2$, then 3.1 up to $\rhob=0.45$, and then 2.7 above. This variation is smaller than that of $\ln(5/\rhob)$, because the nonlinear steepening of the interfaces speeds up the end of the growth. The random projection of the initial perturbation onto the modes close to $k^*$ explains the rest of the scatter.

\subsection{The initial wavelength follows the most amplified wavelength}

Theory also predicts the spacing of the clumps. The question here is how far apart they are when they appear. The most energetic mode does have the wavelength $\lambda^*$ to within \SI{10}{\percent}, with a median ratio of 0.99. The mean wavelength, on the other hand, is about \SI{12}{\percent} shorter. Indeed, the interfaces are already steep at the time of the measurement, and their harmonics shift the spectrum toward large wavenumbers.

The spacing directly shows up in the number of clumps, which increases with $\chi$ (Fig.~\ref{fig:cartes}a). It goes from 4--6 near the threshold to 17 for $\chi=40$. A wavelength $\lambda$ gives about $(L/\lambda)^2$ clumps, that is 2 to 3 near the threshold and 13 for $\lambda^*=4.4$. The order of magnitude is correct. The small excess comes from the quantization of the modes in a domain of only a few wavelengths.

Near the threshold and at low $\rhob$, the scenario changes. There, the clumps appear one after the other, so their number increases with time. The exponent $\alpha$ then becomes negative, which happens in 5 cases out of 47.

\subsection{The area fraction of the clumps equals the mean volume fraction}

We now look at the clumps once they have formed. They concentrate the mass. In median, they contain \SI{93}{\percent} of it (Fig.~\ref{fig:cartes}e). This fraction increases with $\chi$, because a stronger gravity empties the dilute phase more. It also increases with $\rhob$, because the share of mass in the interfaces decreases when the clumps grow.

The area fraction of the clumps equals $\rhob$ to within \SI{4}{\percent} in \SI{80}{\percent} of the cases. The larger deviations reach \SI{25}{\percent}. They concern cases close to the threshold, where the plateaus are not yet saturated at $t=40$. This result has a simple interpretation. The clumps are almost full, and almost all the grains are inside them. Once packed, the grains therefore cover a fraction $\rhob$ of the surface. This is the lever rule of a phase separation. A dilute phase coexists with a dense phase close to $\rho=1$. The Otsu threshold places the interface near $\rho=1/2$, and the mass outside this limit compensates for the density of the plateau, which is slightly below 1. The mean area of a clump is therefore about $\rhob L^2/n$. It increases with $\rhob$ and decreases with $\chi$ (Fig.~\ref{fig:cartes}f).

\begin{figure*}[htb]
\centering
\includegraphics[width=\textwidth]{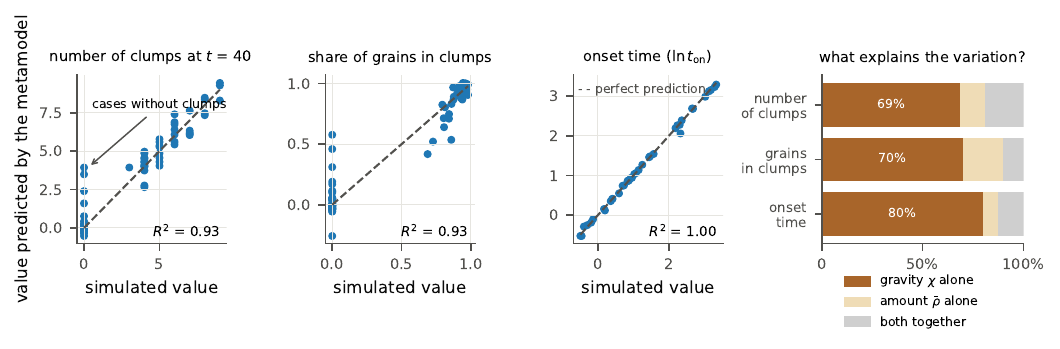}
\caption{The strength of gravity explains most of the results. The first three panels check the metamodel. Each dot compares the value predicted for one simulation, which was not used for the fit, with the simulated value. The cases without clumps form the left column of the first panel. The last panel splits the variation of each result between gravity $\chi$ alone, the amount of grains $\rhob$ alone, and their joint action. These are the first-order Sobol indices.}
\label{fig:meta}
\end{figure*}

\subsection{Coarsening is slow}

The clumps do not stay frozen. After they appear, their number slowly decreases. In median, their number is divided by 1.8 between its maximum and $t=40$. At that time, 3 to 10 clumps remain, whatever the pair $(\chi,\rhob)$ (Fig.~\ref{fig:cartes}b). The pattern therefore loses the memory of its initial wavelength. Yet this coarsening remains slow. Its median exponent is 0.28 on the grid and 0.36 on the hypercube. This slowness agrees with the metastable coarsening known for this model~\cite{dolak2005,potapov2005}, whose speed decreases exponentially with the distance between clumps. A power law therefore gives only a local description of it. A reliable measurement would need a larger domain and a longer time. The areas follow a gamma distribution with a shape parameter close to 3. Finally, the two seeds of the same point differ on average by 1.1 clumps at $t=40$.

\section{Sensitivity analysis}\label{sec:sobol}

The maps show how each observable varies, but not which part of its variance comes from each parameter. Sobol indices answer this question. An index close to 1 means that the parameter explains almost all the variation of the result. An index close to 0 means that it has almost no effect. Their estimation needs tens of thousands of model evaluations. A simulation with clumps takes one to three minutes, so this number is out of reach. We therefore compute them on a metamodel built from the Latin hypercube. A metamodel is a function that is fast to evaluate and that is fitted to imitate the results of the simulator.

The metamodel is a support vector regression with a Gaussian kernel~\cite{smola2004}. Its inputs are $\ln\chi$ and $\rhob$, centered and scaled, and a 5-fold cross-validation sets its hyperparameters. The model of $\ln\ton$ uses only the 40 cases with clumps. Figure~\ref{fig:meta} shows the quality of the fit. The cross-validated coefficient of determination is 0.93 for the number of clumps and for the share of mass in the clumps. It reaches 0.998 for $\ln\ton$.

On this metamodel, the Sobol indices are estimated with $2^{14}$ samples per matrix. The first-order indices use the Saltelli estimator, and the total indices use the Jansen estimator~\cite{saltelli2010}. Table~\ref{tab:sobol} shows that the gravitational number dominates all outputs. The strength of gravity therefore matters more than the amount of grains. The total indices of $\rhob$ are 0.21 to 0.33, while its first-order indices are only 0.07 to 0.20. The mean volume fraction therefore acts mostly through its interaction with $\chi$. This result is expected, because linear stability depends only on $\Lambda=\chi\rhob(1-\rhob)$. However, the indices depend on the explored ranges. Here, $\chi$ varies by a factor of 20, while the product $\rhob(1-\rhob)$ varies only by a factor of 5.

\begin{table}[htb]
  \caption{Sobol indices computed on the metamodel of the Latin hypercube.}
  \label{tab:sobol}
  \begin{ruledtabular}
  \begin{tabular}{lccccc}
    output & $R^2$ & $S_1(\ln\chi)$ & $S_T(\ln\chi)$ & $S_1(\rhob)$ & $S_T(\rhob)$\\
    \midrule
    $n(40)$ & 0.93 & 0.69 & 0.89 & 0.12 & 0.33 \\
    mass fraction & 0.93 & 0.70 & 0.82 & 0.20 & 0.32 \\
    $\ln\ton$ & 1.00 & 0.80 & 0.93 & 0.07 & 0.21 \\
  \end{tabular}
  \end{ruledtabular}
\end{table}

\section{Discussion: three points of view on the formation of solid bodies}\label{sec:discussion}

This work complements two previous studies on the gravitational aggregation of grains~\cite{trivino2025,trivino2026}. The three studies describe the same process at three different scales.

The first study follows each grain individually~\cite{trivino2025}. It uses the discrete element method to simulate a cloud of a few hundred spheres in three dimensions. The grains attract each other through Newtonian gravity and bounce on each other at contacts. There is no gas. The second study works at the scale of the bodies themselves~\cite{trivino2026}. It makes two aggregates of bonded grains collide, and it observes whether they merge or break. The present work sits between the two. It no longer follows the grains but their density, and it describes a dust layer slowed down by the gas. It therefore tells where and when the clumps are born that the two other studies make collapse and then meet.

First, the three studies find the same qualitative result. Aggregation happens only if gravity wins against a mechanism that separates the grains. Here, this mechanism is turbulent spreading, and the threshold is $\Lambda=1+(\pi/L)^2$. In the discrete study, it is the elastic rebound of the grains. Aggregation appears there only above a critical contact dissipation, $\gamma_n\approx5\times10^2$~\cite{trivino2025}. In both cases, the transition from a dispersed state to an aggregated state is sharp.

Second, the two approaches describe opposite regimes of the grain dynamics. In the discrete study, the grains fall freely, and a clump forms in a fraction of the free-fall time $t_{\mathrm{ff}}\sim(G\rho_m)^{-1/2}$, where $\rho_m$ is the mean mass density~\cite{trivino2025}. Here, the gas slows the grains down in a time $t_s$ much shorter than $t_{\mathrm{ff}}$. The growth rate then satisfies $\sigma^*<\Lambda D/\ell^2=4\pi G\rho_m(1-\rhob)\,t_s$. The onset time $\ton\simeq3/\sigma^*$ is therefore at least of the order of $t_{\mathrm{ff}}^2/t_s$, which is much longer than $t_{\mathrm{ff}}$. When the gas slows the grains down, the collapse is slow. The two results therefore do not contradict each other: they bracket the same competition between gravity and dissipation, in the inertial limit on one side and in the damped limit on the other.

The two approaches also agree on the structure of the clumps. In the discrete study, a strong dissipation produces compact clumps, with a fractal dimension close to 3~\cite{trivino2025}. Here, the clumps are almost saturated plateaus, with $\rho\simeq0.99$ at their center. At the continuum scale, volume filling plays the role of the contact repulsion. However, the continuum model assumes compact clumps. It cannot describe the porous and branched clumps that the discrete study obtains at low dissipation.

Finally, coarsening makes the link with the collision study~\cite{trivino2026}. Here, the clumps merge slowly, at the fall velocity set by the gas. These slow encounters correspond to the low impact-stress regime of the collision study, where the two bodies join without breaking. The continuum model cannot represent fragmentation. It therefore remains valid as long as the clumps meet slowly, which is the case for the mergers observed here.

\section{Conclusion}

This work asked whether linear theory predicts the appearance and the spacing of self-gravitating dust clumps, in a model where the grains cannot be packed without limit. To answer this question, a second-order adaptive finite-volume scheme was coupled with an image processing chain and applied to this model. The code converges at second order, reproduces the dispersion relation to within \num{e-4} and conserves mass to machine precision. On the other hand, the adaptive Cartesian mesh brings only about \SI{20}{\percent} at equal cost, because the clumps fill the whole domain.

On the physics side, the marginal stability curve exactly separates the cases with and without clumps. The instability starts at time $\ton\simeq3/\sigma^*$, and the dominant wavelength equals $\lambda^*$ to within \SI{10}{\percent}. The area fraction of the clumps then equals $\rhob$, as predicted by the lever rule. The coarsening that follows is slow, but it erases the memory of the initial wavelength. The gravitational number explains most of the variance, and the mean volume fraction acts mostly through its interaction with it. Put simply, in this model, two numbers are enough to predict when and how far apart the grains gather: the strength of gravity compared with turbulence, and the amount of grains.

These results complement the discrete studies of the fall of a grain cloud and of the collisions between aggregates~\cite{trivino2025,trivino2026}. These quantities are those of an image. They can therefore be compared directly with the density maps of disk simulations, provided that $\chi$ and $\rhob$ are estimated. The same measurements also apply to the amoebae of Keller and Segel, because the equations are identical. An unstructured adaptation would follow the closed interfaces better. A larger domain would make it possible to measure the coarsening law. Finally, adding the rotation of the disk and the effect of the grains on the gas would bring the model closer to real disks.

\end{document}